\documentclass[aps,prl,reprint,groupedaddress]{revtex4-2}

\usepackage[utf8]{inputenc}
\usepackage{latexsym, graphicx} 
\usepackage{amsmath,amsfonts,amssymb}
\usepackage{cancel}
\usepackage{mathrsfs,dsfont}
\usepackage{tensor}
\usepackage{xcolor}
\usepackage{caption,subcaption}
\usepackage[shortlabels]{enumitem}
\usepackage{hyperref}
\hypersetup{
	colorlinks,%
	citecolor=blue,%
	filecolor=blue,%
	linkcolor=blue,%
	urlcolor=blue,
	linktoc=page
}

\newcommand*{\dd}{\mathop{}\!d}
\newcommand{\zbar}{\bar{z}}
\newcommand{\partialbar}{\bar{\partial}}

\begin{document}
	\title{Hydrodynamics of two-dimensional CFTs}
	\author{Kevin Nguyen}
	\email{kevin.nguyen2@ulb.be}
	
	\affiliation{\vspace{0.2cm} Universit\'e Libre de Bruxelles and International Solvay Institutes, ULB-Campus Plaine
		CP231, 1050 Brussels, Belgium}

	\begin{abstract} 
		We demonstrate that the geometric action on a coadjoint orbit of the Virasoro group appropriately describes non-dissipative two-dimensional conformal fluids. While this action had already appeared in the context of AdS$_3$ gravity, the hydrodynamical interpretation given here is new. We use this to argue that the geometric action manifestly controls both sides of the fluid/gravity correspondence, and that the gravitational `hologram' gives an effective hydrodynamical description of the dual CFT.
		As a byproduct, our work sheds light on the nature of the AdS$_3$ reparametrization theory used to effectively compute Virasoro vacuum blocks at large central charge, since the reparametrization mode is now understood as a fluctuation of the fluid velocity. 
	\end{abstract}
	\maketitle
	
	\section{Introduction}
	While the discovery of the AdS/CFT correspondence \cite{Maldacena:1997re,Gubser:1998bc,Witten:1998qj} definitely constitutes a milestone in our quest to understand the nature of gravity on microscopic scales, the continued activity in this field shows that it has yet to deliver its most important lessons. The AdS/CFT correspondence provides a formulation, or perhaps more accurately, a nonperturbative \textit{definition} of quantum gravity with negative cosmological constant in terms of standard conformal field theory (CFT). To a large extent however, it is still unclear how the gravitational `hologram' emerges from this well-defined conformal theory. By now, there is ample evidence that General Relativity should be viewed as an effective field theory (EFT), with the metric field a low-energy degree of freedom which need not be fundamental. In CFT language, the operator dual to the metric is the energy-momentum tensor, which in any Lagrangian theory would arise as a composite field rather than a fundamental one. We believe that this observation should be taken as seriously as possible, and that understanding the emergence of gravity in the context of the AdS/CFT correspondence amounts to the following question:
	\begin{quote}
		\textit{In conformal field theory, what dynamical regime and corresponding effective field theory description looks like gravity with negative cosmological constant?}   
	\end{quote}
	
	We wish to argue that the answer to this question is given by hydrodynamics, which is an effective field theory description of the long-lived modes of a quantum system. In hydrodynamics, a prominent role is played by the components of the energy-momentum tensor and of any other conserved currents, precisely because total energy-momentum conservation and charge conservation prevent them from completely disappearing into the vacuum on any given timescale. They will instead tend to homogeneize over long timescales due to complex and often strong interactions between microscopic degrees of freedom. Note that this is exactly what we are looking for, an effective field theory that keeps track of the energy-momentum tensor on macroscopic scales, which we know is dual to the dynamical metric of the gravitational `hologram'. Adopting a hydrodynamic description of conformal field theory in the appropriate regime, we believe that the correspondence with gravity should become manifest.       
	
	The study of the AdS/CFT correspondence through the lens of hydrodynamics is obviously not new, beginning with the matching of black hole quasinormal modes with poles of thermal response functions \cite{Birmingham:2001pj,Policastro:2002se,Policastro:2002tn}, and culminating in the fluid/gravity correspondence \cite{Baier:2007ix,Bhattacharyya:2007vjd,Bhattacharyya:2008mz,Bhattacharyya:2008xc,Bhattacharyya:2008kq,Rangamani:2009xk,Hubeny:2011hd}. The latter allows, given fluid data up to some given order in the hydro expansion, to generate the corresponding spacetime solution to Einstein's equations with negative cosmological constant. In this paper, we wish to push this correspondence further, to the level of effective actions. The latter have often provided the best way to organise effective field theories, and should allow to make the fluid/gravity correspondence even sharper and insightful.
	
	The construction and classification of effective actions for relativistic hydrodynamics has been the subject of important progress in recent years \cite{Dubovsky:2011sj,Bhattacharya:2012zx,Haehl:2015pja,Haehl:2015uoc,Crossley:2015evo,Haehl:2017zac,Haehl:2018lcu,Campoleoni:2018ltl,Adami:2023fbm}. For the so-called `class L' hydrodynamics, an action principle can be formulated in terms of a reference unnormalized fluid velocity, also known as thermal vector, and its orbit under diffeomorphisms \cite{Haehl:2015pja}. It is worth noting that non-relativistic hydrodynamics had already been formulated and studied by mathematicians using the theory of coadjoint orbits of diffeomorphism groups \cite{Arnold1966SurLG,Arnold1969HamiltonianNO,Ebin,Arnold1998TopologicalMI,Marsden1984SemidirectPA,Marsden1984ReductionAH,Holm1998TheEE,Khesin2020GeometricHA}. (See \cite{Penna:2017vms,Carrillo-Gonzalez:2018wrh,Donnelly:2020xgu} for connections to holography.) Coadjoint orbits of Lie groups are naturally equipped with a symplectic form \cite{Kirillov2004LecturesOT}, such that it is always possible to construct a class of Hamiltonian systems and corresponding action principle which are invariant under the action of the group \cite{Alekseev:1988ce,Alekseev:1990mp,Rai:1989js,Delius:1990pt,Aratyn:1990dj,Barnich:2017jgw,Barnich:2022bni}. The resulting `geometric actions' on coadjoint orbits of diffeomorphism groups would appear to be natural candidates for the description of  `class~L' relativistic hydrodynamics. Returning to the fluid/gravity correspondence, we are thus led to ask the question: 
	\newline
	\begin{quote}
		\textit{Is the fluid/gravity correspondence controlled by a geometric action on the coadjoint orbit of some diffeomorphism group?}   
	\end{quote}
	While we do not have the answer to this question in full generality, the specific setup of AdS$_3$/CFT$_2$ appears ideal to answer it in the affirmative. Indeed, it is already established that the classical solution space of three-dimensional gravity with negative cosmological constant is controlled by the geometric action on coadjoint orbits of the Virasoro group \cite{Barnich:2017jgw}, where the latter is a centrally extended version of the group of diffeomorphisms of the circle (or line). Moreover, this geometric action has been used in the spirit of effective field theory to compute highly nontrivial physical quantities in AdS$_3$/CFT$_2$, such as torus partition functions and Virasoro vacuum blocks at large central charge \cite{Cotler:2018zff,Haehl:2018izb,Haehl:2019eae,Anous:2020vtw,Nguyen:2021jja,Nguyen:2022xsw}. To some extent the Virasoro geometric action seems to control the AdS$_3$/CFT$_2$ correspondence. Its domain of applicability remains unclear however, because the precise nature of this effective field theory, so far called `AdS$_3$ reparametrization theory', is yet to be determined. Following the line of reasoning spelled out above, in this paper we wish to propose that
	\begin{quote}
		\textit{(Claim) The Virasoro geometric action is an effective field theory describing the hydrodynamical regime of CFTs in 2d.}   
	\end{quote}
	If this is correct, then we have found an action that controls the fluid/gravity correspondence in the AdS$_3$/CFT$_2$ setup, which is both manifestly gravitational and hydrodynamical in nature. Here let us note that a similar picture was already established in the context of AdS$_2$ gravity, with the hydrodynamical action for the dual quantum mechanics given by the Schwarzian action \cite{Jensen:2016pah,Mandal:2017thl}. While the relevance of the Virasoro geometric action to AdS$_3$/CFT$_2$ had stemmed from gravitational arguments \cite{Barnich:2017jgw,Cotler:2018zff}, in this letter we unveil its hydrodynamical nature and provide a detailed proof of the above claim. The main novelty of this work therefore lies in the identification of an explicit hydro action for two-dimensional, conformal, non-dissipative fluids. At a more conceptual level, it also unifies existing works dealing with the Virasoro geometric action in the context of AdS$_3$/CFT$_2$, using hydrodynamics as the unifying language.  
	
	We organize this letter as follows. In the next section we describe the geometric action on coadjoint orbits of the Virasoro group and show that it encodes the non-dissipative hydrodynamics of two-dimensional CFTs, namely stress tensor conservation and local entropy conservation, naturally expressed in terms of the orbit of a thermal vector of reference. In the subsequent section we revisit the `reparametrization theory' and explain the nature of the reparametrization mode, which can be understood both as shadow of the stress tensor fluctuation or as fluid velocity fluctuation around thermal equilibrium. In the appendix we illustrate what is meant by hydrodynamical regime in two-dimensional CFT, by discussing real-time response functions close to thermal equilibrium. At leading order in the derivative expansion, we explicitly show that the latter agree with the predictions of a conformal perfect fluid.

	\section{Hydrodynamics from Virasoro geometric action}
	\label{section 3}
	Hydrodynamics is an effective field theory for the long-lived modes of generic interacting systems. Here we will be concerned with neutral fluids, where the only conserved current is the energy-momentum tensor $T_{\mu\nu}$. We will also restrict our attention to two-dimensional conformal fluids. Covering two-dimensional Minkowski space $\mathbb{M}$ with lightcone coordinates
	\begin{equation}
	z=\frac{x+t}{\sqrt{2}}\,, \qquad \zbar=\frac{x-t}{\sqrt{2}}\,,
	\end{equation}
	and corresponding line element
	\begin{equation}
	ds^2=-dt^2+dx^2=2 \dd z \dd\zbar\,,
	\end{equation}
	it is well-known that the components $T\equiv T_{zz}$ and $\bar T\equiv T_{\zbar\zbar}$ provide the two independent degrees of freedom of the stress tensor, subject to the conservation equations
	\begin{equation}
	\label{stress tensor conservation}
	\partialbar T=0=\partial \bar T\,, 
	\end{equation}
	with $\partial\equiv \partial_z$ and $\partialbar \equiv \partial_{\zbar}$. This decomposition into `chiral' sectors renders two-dimensional conformal theories very special. In addition to current conservation, a fundamental ingredient of hydrodynamics is the applicability of a local second law of thermodynamics, expressed by the positive divergence of an entropy current $J_S^\mu$. Here we will restrict our attention to \textit{non-dissipative} fluids, for which the entropy current is conserved \cite{Bhattacharya:2012zx,Haehl:2015pja}. The division of the stress tensor conservation \eqref{stress tensor conservation} into chiral sectors, which will turn into dynamical equations for the fluid variables, suggests that the two components $J_S\equiv J_S^z$ and $\bar J_S\equiv J_S^{\zbar}$ of the entropy current should be separately conserved as well,
	\begin{equation}
	\label{entropy conservation}
	\partial J_S=0=\partialbar \bar J_S\,.
	\end{equation}
	
	We wish to formulate an action principle describing such a two-dimensional, conformal, neutral, non-dissipative fluid. According to the discussion of `class L' fluids, this should be described in terms of orbits of a reference thermal vector field under some diffeomorphism group \cite{Haehl:2015pja}. In addition to the physical spacetime $\mathbb{M}$ equipped with coordinates $(z,\zbar)$, we will therefore also consider a reference spacetime $\tilde{\mathbb{M}}$ diffeomorphic to $\mathbb{M}$ and equipped with coordinates $(F,\bar F)$. Thus $(F(z,\zbar),\bar F(z,\zbar))$ are smooth invertible maps from $\mathbb{M}$ to $\tilde{\mathbb{M}}$. Among them the `holomorphic' maps $F(z), \bar F(\zbar)$, also called conformal transformations, are special in that they map lightrays in $\mathbb{M}$ to lightrays in $\tilde{\mathbb{M}}$. They are the elements of two copies of the diffeomorphism group of the real line, namely
	\begin{equation}
	\begin{aligned}
	F(\,\cdot\,) \in \text{Diff}(\mathbb{R}): \mathbb{R} &\longrightarrow \mathbb{R}\,,\\ 
	z&\longmapsto  F(z) 
	\end{aligned}
	\end{equation}
	and similarly for $\bar F(\zbar)$.
	Because of the chiral split mentioned above, it will be convenient to view generic $F(z,\zbar)$ as one-parameter families of such maps, namely
	\begin{equation}
	\label{one-parameter maps}
	\begin{aligned}
	F(\,\cdot\,,\zbar) \in \text{Diff}(\mathbb{R}): \mathbb{R} &\longrightarrow \mathbb{R}\,, \\
	z&\longmapsto  F(z,\zbar) 
	\end{aligned}
	\end{equation}
	and similarly for $\bar F(z,\zbar)$.
	Typically one would now introduce a timelike thermal vector field $\beta^\mu \partial_\mu$ over $\tilde{\mathbb{M}}$, whose norm would define the inverse temperature $\beta$ via
	\begin{equation}
	\label{inverse temp}
	\beta^2\equiv -\beta_\mu \beta^\mu\,.
	\end{equation} 
	Because of the chiral splitting taking place here, we will instead consider two independent vectors $\beta^F \partial_F$ and $\beta^{\bar F} \partial_{\bar F}$ defined over $\mathbb{R}$, and normalize them independently as
	\begin{equation}
	\beta_F \beta^F=-\frac{1}{2}\beta^2\,, \qquad \beta_{\bar F}\beta^{\bar F}=-\frac{1}{2}\beta^2\,.
	\end{equation}
	The sum of these conditions reproduce \eqref{inverse temp}, and one would say that we are considering equal `chiral' temperatures.  More specifically, we will set
	\begin{equation}
	\beta^F=\beta^{\bar F}=\frac{\beta}{\sqrt{2}}\,, \qquad \beta_F=\beta_{\bar F}=-\frac{\beta}{\sqrt{2}}\,. 
	\end{equation}
	\begin{widetext}
		Their pullback through the one-parameter maps \eqref{one-parameter maps} yields the thermal vector fields in physical spacetime
		\begin{equation}
		\label{thermal fields}
		\begin{split}
		\beta^z(z,\zbar)&=\frac{\beta}{\sqrt{2}}\, (\partial F(z,\zbar))^{-1}\,, \qquad \beta_z(z,\zbar)=-\frac{\beta}{\sqrt{2}}\, \partial F(z,\zbar)\,,\\
		\beta^{\zbar}(z,\zbar)&=\frac{\beta}{\sqrt{2}}\,(\partialbar \bar F(z,\zbar))^{-1}\,, \qquad \beta_{\zbar}(z,\zbar)=-\frac{\beta}{\sqrt{2}}\,\partialbar \bar F(z,\zbar)\,.
		\end{split}
		\end{equation}
		The chiral velocity fields are given by the normalized thermal vector fields, namely
		\begin{equation}
		\label{velocity fields}
		\begin{split}
		u^z(z,\zbar)&=\frac{1}{\sqrt{2}}\, (\partial F(z,\zbar))^{-1}\,, \qquad u_z(z,\zbar)=-\frac{1}{\sqrt{2}}\, \partial F(z,\zbar)\,,\\
		u^{\zbar}(z,\zbar)&=\frac{1}{\sqrt{2}}\,(\partialbar \bar F(z,\zbar))^{-1}\,, \qquad u_{\zbar}(z,\zbar)=-\frac{1}{\sqrt{2}}\,\partialbar \bar F(z,\zbar)\,.
		\end{split}
		\end{equation}
	\end{widetext}
	The quantities $\left(u^z(z,\zbar),u^{\zbar}(z,\zbar),\beta\right)$ make up the basic physical constituents of the hydrodynamical theory we wish to formulate. To zeroth order in the hydro expansion, we recall that we should have the constitutive relation of a perfect two-dimensional conformal fluid, namely
	\begin{equation}
	\label{perfect fluid}
	T_{\mu\nu}=(2 u_\mu u_\nu+\eta_{\mu\nu})\, \varepsilon+O(\partial u)\,,
	\end{equation}
	with $\varepsilon$ the energy density at inverse temperature $\beta$,
	\begin{equation}
	\varepsilon=\langle T_{tt} \rangle_\beta=\frac{1}{2} \left(\langle T \rangle_\beta+\langle \bar T \rangle_\beta \right)=\langle T \rangle_\beta\,.
	\end{equation}
	In lightcone coordinates $(z,\zbar)$ and upon using \eqref{velocity fields}, the constitutive relation \eqref{perfect fluid} becomes
	\begin{equation}
	\label{perfect fluid 2}
	T=2\varepsilon (u_z)^2+O(\partial u)= \varepsilon\, (\partial F)^2+O(\partial^2 F)\,.
	\end{equation}
	The identity map $F(z,\zbar)=z$ characterizes the fluid flow of reference, i.e., thermal equilibrium.
	
	Now we come to the problem of formulating an action principle encoding all the above information. Because of the chiral splitting of the stress tensor, we expect the action to split accordingly. As anticipated in the introduction, a natural candidate for the hydrodynamical action is the geometric action on a coadjoint orbit of the Virasoro group, given by \cite{Alekseev:1988ce}
	\begin{widetext}
		\begin{equation}
		\label{unit cylinder action}
		S[F]=\int \dd z \dd \zbar \left(-b_0[F]\, \partial F\, \bar \partial F +\frac{c}{48\pi} \frac{\bar \partial F}{\partial F} \left[\frac{\partial^3 F}{\partial F}-2\left(\frac{\partial^2 F}{\partial F} \right)^2 \right]\right)\,,
		\end{equation}
	\end{widetext}
	where $F$ satisfies the periodicity condition $F(z+2\pi,\zbar)=F(z,\zbar)+2\pi$ (we will come back to the meaning of this condition). The action for the other chiral sector is obtained by replacing $F \mapsto \bar F$ together with $\partial \leftrightarrow \partialbar$. \newline
	The resulting equation of motion \textit{is} the conservation of the stress tensor
	\begin{equation}
	\bar \partial T=0\,, 
	\end{equation}
	where the constitutive expression of the latter in terms of $F$ is given by
	\begin{equation}
	\label{T solution}
	T\equiv b_0[F]\, (\partial F)^2 -\frac{c}{24\pi} S[F]\,,
	\end{equation}
	with the Schwarzian derivative
	\begin{equation}
	S[F]\equiv \frac{\partial^3 F}{\partial F}-\frac{3}{2}\left(\frac{\partial^2 F}{\partial F}\right)^2\,.
	\end{equation}
	The functional $b_0[F]$ labels a coadjoint orbit. We argue that the appropriate $b_0$ corresponds to the orbit containing the lowest energy state on the cylinder \cite{Barnich:2014zoa}, namely
	\begin{equation}
	\label{b0}
	b_0=-\frac{c}{48\pi}\,.
	\end{equation}
	With that constant value and  using \eqref{T solution}, the fluid solution of reference given by the identity map is characterized by 
	\begin{equation}
	F_0(z,\zbar)=z \quad \longrightarrow \quad T_0=b_0=-\frac{c}{48\pi}\,.
	\end{equation}
	This is indeed the correct expectation value of the stress tensor $\langle T \rangle_\beta$ of a two-dimensional CFT at inverse temperature $\beta=2\pi$. Note that having $\beta=2\pi$ is simply a choice of units as $\beta$ is the only dimensionful parameter in the problem. We can however restore an arbitrary $\beta$ if we like by imposing the modified periodicity $F(z+\beta,\zbar)=F(z,\zbar)+2\pi$, such that the reference fluid solution is now given by
	\begin{equation}
	F_\beta(z,\zbar)=\frac{2\pi}{\beta}\, z \quad \longrightarrow \quad  \langle T\rangle_\beta= -\frac{c}{48\pi}\left(\frac{2\pi}{\beta}\right)^2\,. 
	\end{equation}
	Obviously this modified periodicity is that of the reference euclidean cylinder where thermal field theory is naturally defined. In what follows we will keep $\beta=2\pi$. We can also easily check that the stress tensor \eqref{T solution} indeed encodes the expected constitutive relation \eqref{perfect fluid 2},
	\begin{equation}
	T=b_0\, (\partial F)^2+O(\partial^2 F)=\langle T\rangle_\beta\, (\partial F)^2+O(\partial^2 F)\,.
	\end{equation}
	
	In addition, due to the particular value \eqref{b0}, the geometric action \eqref{unit cylinder action} enjoys $\operatorname{PSL}(2,\mathbb{R})$ symmetry which is the stabilizer of the corresponding exceptional orbit \cite{Alekseev:1988ce}. In particular, the infinitesimal symmetry
	\begin{equation}
	\delta F(z,\zbar)=\varepsilon(z,\zbar) F(z,\zbar)\,,
	\end{equation}
	is associated to the conserved current
	\begin{equation}
	J_S=\frac{c}{24\sqrt{2}\pi} \left(\frac{\partial^2 \partialbar F}{(\partial F)^2}-\frac{\partial^2 F \partial \partialbar F}{(\partial F)^3}+ \partialbar F\right)\,.
	\end{equation}
	Besides satisfying \eqref{entropy conservation} onshell, we can explicitly see that it satisfies the offshell relation
	\begin{equation}
	\partial J_S=-\frac{1}{\sqrt{2}}(\partial F)^{-1}\, \partialbar T=-\beta^z \partialbar T\,,    
	\end{equation}
	where in the last equality we used the thermal vector field \eqref{thermal fields}. This a chiral version of the \textit{adiabaticity condition} satisfied by the conserved entropy current of a neutral fluid \cite{Haehl:2015pja},
	\begin{equation}
	\partial_\mu J_S^\mu=-\beta_\mu \partial_\nu T^{\mu\nu}\,.
	\end{equation}
	It is very interesting to note that the $\operatorname{PSL}(2,\mathbb{R})$ symmetry underlying this entropy current comes for free in this model, which is to be contrasted with the general discussion found in \cite{Haehl:2015pja}. Note that it would also be interesting to understand the nature of the two other $\operatorname{PSL}(2,\mathbb{R})$ currents found in \cite{Alekseev:1988ce}.
	As anticipated, the geometric action on the first exceptional coadjoint orbit of the Virasoro group, characterized by the constant covector \eqref{b0}, encodes all the information regarding the hydrodynamics of a neutral, non-dissipative, conformal, two-dimensional fluid. Since it was already shown to control the AdS$_3$ gravitational hologram \cite{Barnich:2017jgw,Cotler:2018zff}, this completes the proof that it actually controls both sides of the fluid/gravity correspondence. We believe that this illuminates the AdS$_3$/CFT$_2$ correspondence by providing a unique language describing both sides of the duality at once!
	
	\section{Reparametrization theory revisited}
	\label{section 4}
	As explained in the introduction, the Virasoro geometric action \eqref{unit cylinder action} has also been used as the basis for computing Virasoro vacuum blocks at large central charge \cite{Cotler:2018zff,Haehl:2018izb,Haehl:2019eae,Anous:2020vtw,Nguyen:2021jja,Nguyen:2022xsw}. This effective field theory of a new kind was termed `reparametrization theory', and its rules invented `as we go' such as to reproduce known results, until they could be derived from first principles by the author \cite{Nguyen:2021jja,Nguyen:2022xsw}. Here we wish to give another look at this reparametrization theory, armed with the new understanding of the Virasoro geometric action as describing the hydrodynamical regime of two-dimensional CFTs. 
	
\begin{figure}
		\centering
		\includegraphics[scale=0.4]{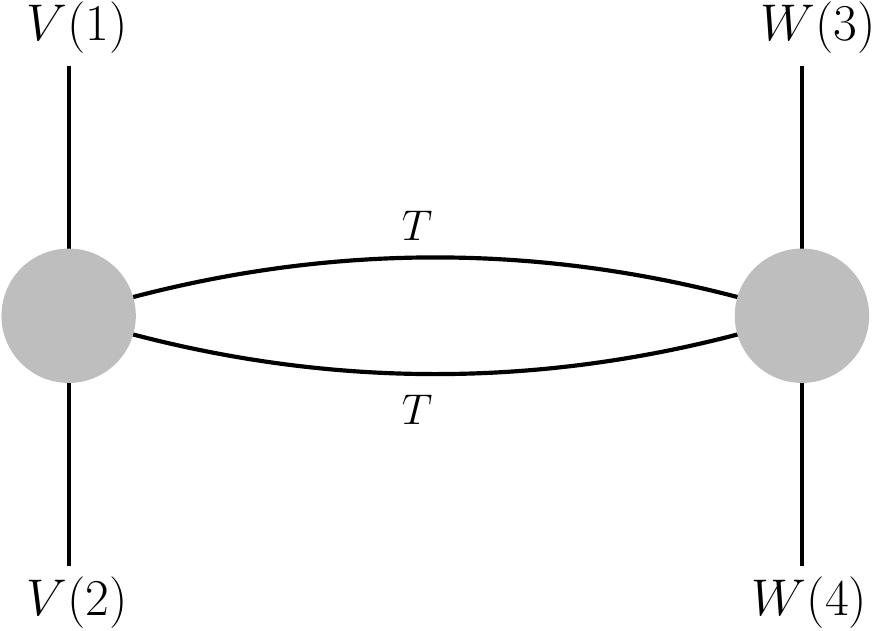}
		\caption{Feynman diagram corresponding to stress tensor exchanges between two pairs of identical operators, borrowed from \cite{Nguyen:2021jja}. The exchange of the stress tensor operator is effectively computed by exchanging the reparametrization mode $\epsilon(z,\zbar)$ using a set of effective Feynman rules.}
		\label{fig:Feynman}
	\end{figure}

	To set up the computation of Virasoro blocks, one takes the Virasoro geometric action \eqref{unit cylinder action} and expands it in terms of a fluctuation $\epsilon(z,\zbar)$ around the identity map $F=z$\,,
	\begin{equation}
	\label{F expansion}
	F=z+\epsilon(z,\zbar)+O(\epsilon^2)\,,
	\end{equation}
	such that the action \eqref{unit cylinder action} becomes
	\begin{widetext}
		\begin{equation}
		S[\epsilon]=\frac{c}{48\pi}\int \dd z \dd \zbar \left( \left(\frac{2\pi}{\beta}\right)^2(\partialbar \epsilon+\partial \epsilon\, \partialbar \epsilon) + \partial^3 \epsilon\, \partialbar \epsilon +O(\epsilon^3)\right)\,.
		\end{equation}
	\end{widetext}
	Here we have reinstated the temperature dependence~$\beta$.
	This action is then treated exactly like in standard perturbation theory, building the propagator of the `reparametrization mode'~$\epsilon$ by inverting the kinetic term, and treating higher-order terms $O(\epsilon^3)$ as interaction vertices \cite{Haehl:2019eae,Anous:2020vtw,Nguyen:2021jja}. In particular, taking the zero-temperature limit $\beta \to \infty$, the propagator simply reads \cite{Haehl:2019eae,Anous:2020vtw,Nguyen:2021jja}
	\begin{equation}
	\langle \epsilon(z_1,\zbar_1) \epsilon(z_2,\zbar_2) \rangle =\frac{6}{c} z_{12}^2 \ln z_{12}\,.
	\end{equation}
	To couple $\epsilon$ to other primary operators, one introduces bilocal vertex operators which are reparametrized two-point functions of the operators of interest.
	Because of the appearance of the factor $c^{-1}$, the resulting perturbation theory is naturally suited to the regime of large central charge $c \gg 1$. One can then compute the Virasoro vacuum block contribution to arbitrary four-point functions of the form $\langle V V W W \rangle$ from Feynman diagrams like the one displayed in Figure~\ref{fig:Feynman}. For details the reader should consult \cite{Haehl:2019eae,Anous:2020vtw,Nguyen:2021jja,Nguyen:2022xsw}.
	
	In this section we simply wish to comment on the nature of the `reparametrization mode'~$\epsilon$, and connect it in particular to fluid variables. Combining \eqref{velocity fields} and \eqref{F expansion}, we find that the fluctuation of the velocity field around thermal equilibrium is simply given by
	\begin{equation}
	\delta u^z=\delta u_z=-\frac{1}{\sqrt{2}}\, \partial \epsilon\,.
	\end{equation}
	On the other hand, combining \eqref{T solution} and \eqref{F expansion} yields the stress tensor fluctuation
	\begin{equation}
	\delta T=-\frac{c}{24\pi}\left(\left(\frac{2\pi}{\beta}\right)^2 \partial +\partial^3 \right)\epsilon\,.
	\end{equation}
	This just happens to be the \textit{shadow transform} at finite temperature, as originally defined in \cite{Ferrara:1972uq}. To make this manifest, we write
	\begin{widetext}
		\begin{align}
		\delta T(z)&= - \frac{c}{24\pi}\left[\left(\frac{2\pi}{\beta}\right)^2 \partial_z +\partial_z^3 \right] \int d^2w\, \delta^{(2)}(z-w)\, \epsilon(w,\bar w) =- \frac{c}{24\pi^2}\left[\left(\frac{2\pi}{\beta}\right)^2 \partial_z +\partial_z^3 \right] \int d^2w\, \partial_{\bar w}\left(\frac{1}{w-z} \right)\, \epsilon(w,\bar w)\nonumber \\
		\label{shadow}
		&=-\frac{c}{24\pi^2} \int d^2w\, \left[\left(\frac{2\pi}{\beta}\right)^2 \partial_z +\partial_z^3 \right]\left(\frac{1}{z-w} \right)\, \partialbar \epsilon(w,\bar w) =\frac{c}{4\pi^2} \left(\frac{\pi}{\beta} \right)^4 \int d^2w\, \frac{1}{\sin^4 \frac{\pi}{\beta}(z-w)}\, \partialbar \epsilon(w,\bar w)\,, 
		\end{align}
	\end{widetext}
	where in the second equality we used the distributional identity from complex analysis, $\partial_{\bar z}(1/z)=\pi \delta^{(2)}(z,\zbar)$.
	In the last expression of \eqref{shadow} we recognize the connected stress tensor two-point function \cite{Nguyen:2021jja}
	\begin{equation}
	\begin{split}
	&\langle T(z) T(w) \rangle=\left(\frac{\pi}{\beta} \right)^4 \frac{c}{2 \sin^4 \frac{\pi}{\beta} (z-w) }\\
	&=-\frac{c}{12}\left[\left(\frac{2\pi}{\beta}\right)^2 \partial_z +\partial_z^3 \right]\left(\frac{1}{z-w} \right)+\text{regular}\,,
	\end{split}
	\end{equation}
	such that we can identify $\partialbar \epsilon$ in the last line of \eqref{shadow} as the shadow of the stress tensor fluctuation. Thus the reparametrization mode $\epsilon$, used as dynamical field variable in the reparametrization theory, can be rightfully interpreted as shadow of the stress tensor as first suggested in \cite{Haehl:2018izb,Haehl:2019eae}, or as fluid variable as demonstrated in this work. 
	
	\section{Conclusions}
	In this work, we demonstrated that the well-known geometric action on a coadjoint orbit of the Virasoro group is the appropriate hydro action for non-dissipative, two-dimensional conformal fluids. We showed that this action encodes the conservation of the stress tensor as well as the conservation of an entropy current. At leading order in the hydro expansion, it reproduces the constitutive relations of a perfect conformal fluid. The identification of the hydro action which appropriately describes this class of fluids is the main result of this work. Importantly, it illuminates earlier works dealing with the Virasoro geometric action in the context of AdS$_3$ gravity and holography, by unifying them within the language of hydrodynamics. In particular, we explicitly revisited the `reparametrization theory' used to compute vacuum blocks at large central charge, and framed it as a hydro computation where fluctuations of the fluid velocity around thermal equilibrium are considered and quantized. These findings should motivate a more systematic EFT approach to holography and quantum gravity, where fluid dynamics acts as the macroscopic manifestation of the quantum gravitational dynamics of higher-dimensional spacetimes. 
	
	\section{Acknowledgments}
	I thank David Ramirez and Felix Haehl for insightful discussions. I thank the organizers of the workshop ``Gravity - New perspectives from strings and higher dimensions" and Centro de Ciencias de Benasque Pedro Pascual where part of this project was initiated. This work is supported by a Postdoctoral Research Fellowship granted by the F.R.S.-FNRS (Belgium).
	
	\section{Appendix: Hydrodynamics of CFT$_2$}
	
	In this section we show that retarded thermal Green's functions of a two-dimensional CFT agree with the predictions corresponding to a perfect conformal fluid. This allows to illustrate what is meant by the hydrodynamical regime of two-dimensional CFT. These computations have been provided to me by David Ramirez, to whom I am much indebted. 
	
	\subsection*{Hydrodynamical predictions}
	Following the Kadanoff--Martin approach \cite{kadanoff-martin}, we can predict the form of the retarded thermal
	Green's functions by solving an initial value problem for hydrodynamic fluctuations
	around an equilibrium state, and comparing it to the result obtained from linear
	response theory. For a modern overview of hydrodynamics using this approach, see
	\cite{Kovtun:2012rj}. To zeroth order in the derivative expansion, the constitutive relation for the stress tensor of the fluid is given by
	\begin{align}
	\label{eq:ideal-T}
	T^{\mu\nu} ={}& \varepsilon\, u^\mu u^\nu + p\, \Delta^{\mu\nu} = (\varepsilon + p)\, u^\mu u^\nu + p\, \eta^{\mu\nu}\,,
	\end{align}
	where $\varepsilon$ is the energy density, $p$ the pressure, $u^\mu$ the fluid velocity normalized to unity, and $\Delta^{\mu\nu} = \eta^{\mu\nu} + u^\mu u^\mu$ the projector onto the space tranverse to this velocity. We will be considering a fluid moving in two-dimensional Minkowski space, with coordinates $x^\mu=(t,x)$. For a fluid at
	rest ($\bar u \equiv \partial_t$) with constant energy density
	$\bar \varepsilon$ and pressure $\bar p$, we
	simply have
	\begin{align}
	\bar T^{\mu\nu} =
	\begin{pmatrix}
	\bar \varepsilon & 0 \\ 0 & \bar p
	\end{pmatrix}\,.
	\end{align}
	
	We now look at fluctuations around this equilibrium
	configuration. Therefore, we write
	$\varepsilon = \bar \varepsilon + \delta \varepsilon$,
	$u^\mu = \bar u^\mu +\delta u^\mu$, and $p = \bar p + \delta p$, and solve
	$\partial_\mu T^{\mu\nu} = 0$ to linear order in terms of
	initial data. We note that the normalization $u^2=-1$ implies $\delta u^t=0$.  The
	linearized stress tensor is then given by
	\begin{align}
	\label{eq:deltaT}
	\delta T^{\mu\nu} =
	\begin{pmatrix}
	\delta \varepsilon & \bar w\, \delta u^x \\ \bar w\, \delta u^x & \delta p
	\end{pmatrix}\, ,
	\end{align}
	where we have introduced the (equilibrium) enthalpy
	$\bar w = \bar \varepsilon + \bar p$. Note that the momentum
	density fluctuation is related to the velocity fluctuation via
	$\delta \pi^x= \bar w\, \delta u^x$. Stress tensor conservation then yields two equations
	\begin{equation}
	\label{eq:dteps}
	\begin{split}
	0 ={}& \partial_t \delta \varepsilon + \partial_x \delta \pi^x\,, \\
	0 ={}& \partial_t \delta \pi^x + \partial_x \delta p\, .
	\end{split}
	\end{equation}
	To find a closed set of equations for $\delta \varepsilon$ and
	$\delta u^x$, we introduce the speed of sound $c_s$ at equilibrium, 
	\begin{align}
	\delta p = \left( \frac{\partial p}{\partial \varepsilon}\right) \delta \varepsilon \equiv c_s^2\, \delta \varepsilon\, .
	\end{align}
	Plugging this into \eqref{eq:dteps} one can find standard wave
	equations for $\delta \varepsilon$ and $\delta \pi^x$, showing that
	the normal modes of the system are unattenuated sound waves
	traveling at the speed of sound $c_s$. However, let's instead Laplace
	transform in time (and Fourier transform in space) which is better suited to analyze the initial value problem. With the convention
	\begin{equation}
	f(z,k) \equiv \int_0^\infty d t\, \int_{-\infty}^\infty d x\, e^{i (z t - k x)} f(t,x)\,,
	\end{equation}
	doing so yields
	\begin{align}
	\begin{pmatrix}
	\delta \varepsilon (t=0, k) \\ \delta \pi^x(t=0, k) 
	\end{pmatrix} =
	i \begin{pmatrix}
	-z & k \\ c_s^2 k & -z
	\end{pmatrix}
	\begin{pmatrix}
	\delta \varepsilon (z, k) \\ \delta \pi^x(z, k) 
	\end{pmatrix}\, .
	\end{align}
	Denoting the matrix here by $M_{ab}$ and the hydrodynamic fluctuations
	collectively as $\delta \phi^a$, we simply invert the matrix to solve
	for $\delta \phi^a(z,k)$,
	\begin{align}
	\label{eq:hydro-IVP}
	\delta \phi^a (z, k) = (M^{-1})\indices{^a_b}\, \delta \phi^b(t=0,k)\, .
	\end{align}
	
	To extract the Green's function, we compare this result to what we
	find via linear response theory. The fundamental result in linear response theory
	says that the change in an expectation value
	$\langle \phi_a\rangle$ upon turning on a small source $\delta J^a$
	is determined by the equilibrium Green's function via
	\begin{equation}
	\begin{split}
	\label{eq:lin-res}
	&\delta \langle\phi_a(t,x)\rangle\\
	&=-\!\int_{-\infty}^\infty \mathrm{d} t'\! \int_{-\infty}^\infty d x' G^R_{ab}(t-t',x- x') \delta J^b(t',x')\,,
	\end{split}
	\end{equation}
	where the retarded Green's function is defined as
	\begin{align}
	G^R_{ab}(t\!-\!t',x\!-\!x') =\!-i \theta(t\!-\!t') \langle [ \phi_a(t,x), \phi_b(t',x')] \rangle .
	\end{align}
	We apply this technology to the following scenario: we turn on an infinitessimal, static source $\delta J^a$ in the far past, let the system react up until $t=0$ where we turn the source off, and then measure the resulting relaxation back to equilibrium. In other words, we take
	\begin{align}
	\delta J^a(t,k) =
	\begin{cases}
	\delta J^a(k)\, e^{\epsilon t} & t \leq 0 \\ 0 & t>0
	\end{cases}\,,
	\end{align}
	where $\epsilon$ will be some positive number.
	Using \eqref{eq:lin-res}, the expectation value at $t=0$ is then given by
	\begin{widetext}
		\begin{equation}
		\label{delta O(0)}
		\delta \langle \phi_a (t=0,k)\rangle =- \int_{-\infty}^0 \mathrm{d} t \, e^{\epsilon t} G^R_{ab}({-} t, k)\, \delta J^b(k) = - G^R_{ab}(z=i \epsilon, k)\, \delta J^b(k) \equiv \chi_{ab}(k)\, \delta J^b(k)\, ,
		\end{equation}
	\end{widetext}
	while for $t>0$, 
	\begin{align}
	\delta \langle \phi_a (t,k)\rangle = {-} \delta J^b(k)\int_{-\infty}^0 \mathrm{d} t' \, e^{\epsilon t'} G^R_{ab}(t- t', k) \, .
	\end{align}
	Laplace transforming this second expression, one finds after some
	manipulation,
	\begin{equation}
	\begin{split}
	&\delta \langle \phi_a(z,k)\rangle = \delta J^b(k) \int \frac{\mathrm{d} \omega}{2\pi} \frac{G^R_{ab}(\omega,k)}{(\omega-i \epsilon) (\omega -z)}\\
	&= - \frac{G^R_{ab}(z,k) - G^R_{ab}(z=i\epsilon,k)}{i(z- i \epsilon)}\, \delta J^b(k)\, .
	\end{split}
	\end{equation}
	Here we've closed the integration contour in the upper half plane,
	using the analyticity of $G^R(\omega)$ there that follows from
	causality. Finally trading $\delta J^a(k)$ for
	$\delta \langle \phi^a(t=0,k)\rangle$ using \eqref{delta O(0)}, we conclude
	\begin{equation}
	\label{eq:lin-res-IVP}
	\begin{split}
	&\delta \langle \phi_a(z,k)\rangle\\
	&= - \frac{\left[G^R(z,k)\chi(k)^{-1} + 1\right]\indices{_a^b}}{i (z-i\epsilon)}\, \delta \langle\phi_b(t=0,k)\rangle\, .
	\end{split}
	\end{equation}
	Identifying \eqref{eq:hydro-IVP} and \eqref{eq:lin-res-IVP}, we can now write
	\begin{equation}
	\begin{split}
	G^R(z,k)& = -[i (z-i\epsilon) M^{-1} + 1]\, \chi(k)\\
	&=\frac{1}{z^2 -c_s^2 k^2}
	\begin{pmatrix}
	c_s^2 k^2  & z k \\ c_s^2 k z & c_s^2 k^2
	\end{pmatrix}
	\chi(k)\, .
	\end{split}
	\end{equation}
	The last ingredient we need is the susceptibility matrix
	$\chi(k\to 0)$,
	\begin{align}
	\chi(k\to 0) = 
	\begin{pmatrix}
	\bar w/c_s^{2}  & 0 \\ 0 & \bar w
	\end{pmatrix}\,.
	\end{align}
	Putting this altogether, 
	we can write down the hydro prediction for the retarded Green's function of the energy density at small momentum $k$, 
	\begin{align}
	\label{energy two-point function}
	G^R_{\varepsilon,\varepsilon}(\omega,k) ={}& \frac{\bar w k^2}{\omega^2 -c_s^2 k^2}\,, \qquad (k\to 0)\,.
	\end{align}
	
	\subsection*{Agreement with CFT$_2$ result}
	Hydrodynamics gave a prediction for the thermal retarded Green's function of the stress tensor to lowest order in momentum $k$ in a generic two-dimensional theory. Let us check that this prediction indeed applies to two-dimensional CFT at finite temperature. 
	
	First we note that a conformal fluid at rest 
	satisfies $\bar \varepsilon = \bar p$ from tracelessness of $T^{\mu\nu}$,
	such that $\bar w = \bar \varepsilon +\bar p = 2 \bar \varepsilon$ and
	$c_s^2 = \partial p/ \partial \varepsilon =1$. Less trivially, the existence of Virasoro symmetry in two dimensions implies that the
	$T_{tt}$ two-point function is completely fixed in terms of the
	energy density $\bar \varepsilon$ \cite{Ramirez:2020qer},
	\begin{equation}
	\begin{split}
	G^R_{T_{tt},T_{tt}}(\omega,k) &= \frac{\omega^2}{\omega^2 - k^2 } \left(2 \bar \varepsilon + \frac{c \omega^2}{12 \pi} \right)\\
	&= \frac{\bar w k^2}{\omega^2 -k^2} + \bar w + \frac{c}{12 \pi} \frac{\omega^4}{\omega^2 -k^2}\, .
	\end{split}
	\end{equation}
	So up to a constant and higher order terms, which are not captured by the leading
	hydrodynamic analysis presented above, we see that the CFT$_2$ result reproduces
	the hydro expectation \eqref{energy two-point function}.

	\bibliography{bibl}
	\bibliographystyle{JHEP}  
	
\end{document}